# Subwavelength particles in an inhomogeneous light field: Optical forces associated with the spin and orbital energy flows

**A Ya Bekshaev**

I I Mechnikov National University, Dvorianska 2, Odessa 65082, Ukraine

E-mail: bekshaev@onu.edu.ua

**Abstract**
We analyze the ponderomotive action experienced by a small spherical particle immersed in an optical field, in relation to the internal energy flows (optical currents) and their spin and orbital constituents. The problem is studied analytically, based on the dipole model, and numerically. Three sources of the field mechanical action – energy density gradient and the orbital and spin parts of the energy flow – differ by the ponderomotive mechanism, and their physical nature manifests itself in the optical force dependence on the particle radius $a$. If $a \ll \lambda$ (the radiation wavelength), the optical force behaves as $a^{\nu}$ and integer $\nu$ can be used to classify the sources of the mechanical action. This classification correlates with the multipole representation of the field – particle interaction: The gradient force and the orbital-momentum force appear due to the electric or magnetic dipole moments per se, the spin-momentum force emerges due to interaction between the electric and magnetic dipoles or between the dipole and quadrupole moments (if the particle is polarisable electrically but not magnetically or vice versa). In principle, the spin and orbital currents can be measured separately by the probe particle motion, employing the special choice of particles with necessary magnetic and/or electric properties.

## 1. Introduction

Our subject will be a monochromatic electromagnetic field with electric and magnetic vectors $\mathrm{Re}\left[\mathbf{E}\exp\left(-i\omega t\right)\right]$, $\mathrm{Re}\left[\mathbf{H}\exp\left(-i\omega t\right)\right]$. In general situations, the field amplitudes $\mathbf{E}$ and $\mathbf{H}$ are inhomogeneously distributed over the 3D space parameterized by the Cartesian frame with the unit vectors $\mathbf{e}_x$, $\mathbf{e}_y$, $\mathbf{e}_z$ and the radius-vector $\mathbf{R} = x\mathbf{e}_x + y\mathbf{e}_y + z\mathbf{e}_z = \mathbf{r} + z\mathbf{e}_z$ (explicit separation of the transverse and longitudinal parts of **R** in the latter representation is natural for paraxial fields [1] but even in cases far from paraxiality there usually exists a certain physically highlighted longitudinal direction which we associate with axis *z*). Such a field possesses two important time-average characteristics [2]: the energy density, which in a homogeneous dielectric medium with permittivity $\varepsilon$ and permeability $\mu$ is described by relation

$$w = w_e + w_m = \frac{g}{2}\left(\varepsilon \left|\mathbf{E}\right|^2 + \mu \left|\mathbf{H}\right|^2\right), \tag{1}$$

(the electric $w_e$ and magnetic $w_m$ energy densities equal to the first and second summands in the right-hand side, respectively), and the energy flow density expressed by the Poynting vector **S** and proportional to the field momentum density **p**,

$$\mathbf{S} = c^2\mathbf{p} = gc\,\mathrm{Re}\left(\mathbf{E}^* \times \mathbf{H}\right). \tag{2}$$

Here $g = \left(8\pi\right)^{-1}$ in the Gaussian system of units, *c* is the velocity of light in vacuum, and fornula (2) represents the kinetic (Abraham) momentum density of the electromagnetic field [3,4].

An important quality of the field momentum is its complex structure: it can be subdivided into the physically different spin momentum density (SMD) $\mathbf{p}_S$ and orbital momentum density (OMD) $\mathbf{p}_O$ which reflect the special energy flow features associated with the spin and spatial degrees of freedom of light [4–8],

$$\mathbf{p} = \mathbf{p}_S + \mathbf{p}_O \,. \tag{3}$$

In a homogeneous non-absorbing medium, the separate terms of the field momentum decomposition (3) read [9,10]

$$\mathbf{p}_S = \mathbf{p}_S^e + \mathbf{p}_S^m, \quad \mathbf{p}_S^e = \frac{g}{4\omega}\mathrm{Im}\,\nabla \times \left(\frac{1}{\mu}\mathbf{E}^* \times \mathbf{E}\right), \quad \mathbf{p}_S^m = \frac{g}{4\omega}\mathrm{Im}\,\nabla \times \left(\frac{1}{\varepsilon}\mathbf{H}^* \times \mathbf{H}\right), \tag{4}$$

$$\mathbf{p}_O = \mathbf{p}_O^e + \mathbf{p}_O^m, \quad \mathbf{p}_O^e = \frac{g}{2\omega}\mathrm{Im}\left[\frac{1}{\mu}\mathbf{E}^* \cdot \left(\nabla\right)\mathbf{E}\right], \quad \mathbf{p}_O^m = \frac{g}{2\omega}\mathrm{Im}\left[\frac{1}{\varepsilon}\mathbf{H}^* \cdot \left(\nabla\right)\mathbf{H}\right]. \tag{5}$$

In (4) and (5), the analytically identical contributions of the electric **E** and magnetic **H** fields (‘electric-magnetic democracy’ [5]) are explicitly exposed, and $\mathbf{C}^* \cdot \left(\nabla\right)\mathbf{C}$ is the invariant notation for the vector operation that in Cartesian components reads [6,7]

$$\mathbf{C}^* \cdot \left(\nabla\right)\mathbf{C} = C_x^*\nabla C_x + C_y^*\nabla C_y + C_z^*\nabla C_z \,.$$

The total field momentum (3) contains the ‘trivial’ longitudinal component directed along the (nominal) propagation direction *z* and characterizing the ‘overall’ energy flow of the considered field; all the residual contributions express internal energy flows (optical currents), which usually constitute the main research interest. Representation (3) – (5) highlights the physical difference and peculiar characters of the spin and spatial degrees of freedom and provides efficient tools for analysis of their interactions and interrelations.

The OMD (5) expresses the local expectation value of the field momentum operator $-i\nabla$, and its internal part characterizes the spatial energy redistribution within an optical field in the course of its evolution. For example, in a paraxial beam propagating along the longitudinal axis *z*, the transverse OMD satisfies the ‘continuity equation’ [7]

$$\frac{\partial w}{\partial z} = -c\left(\nabla_\perp \mathbf{p}_{O\perp}\right) \tag{6}$$

where the subscript $\perp$ symbolizes the transverse part of a vector, $\mathbf{p}_\perp = \mathbf{e}_x p_x + \mathbf{e}_y p_y$. In contrast, the SMD (4) is not associated with any real spatial energy motion: it originates from the “hidden” rotation of the field vectors and in paraxial beams is directly linked to inhomogeneity of the local degree of circular polarization, usually expressed via the ‘third Stokes parameter’ $s_3(\mathbf{r})$ [2,7]. The SMD is oriented along the constant-level lines of function $s_3(\mathbf{r})$, is always divergenceless, $\nabla_\perp \mathbf{p}_{S\perp} = \nabla \mathbf{p}_S = 0$, and thus gives no contribution to the total field momentum [7,8].

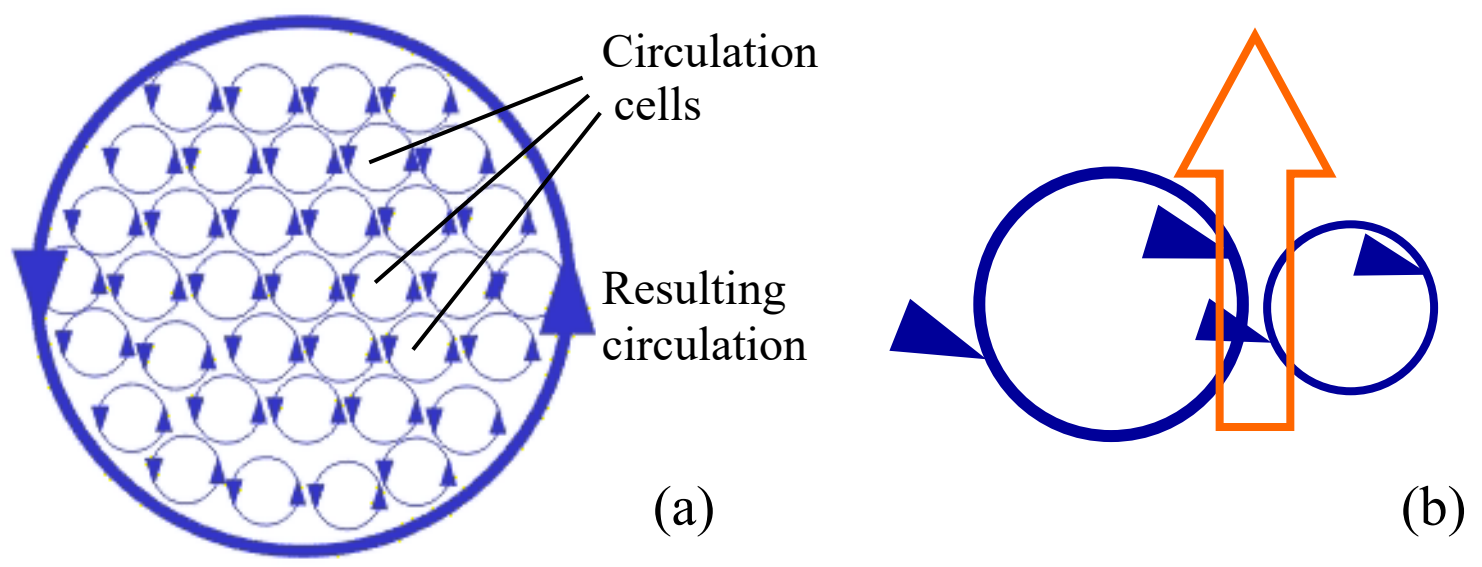

**Figure 1**. Model of the spin flow formation in a beam with circular polarization: (a) homogeneous beam with abrupt boundary, (b) emergence of the macroscopic spin flow (vertical arrow) in case of the beam inhomogeneity (difference of the cell circulations is denoted by the difference in their sizes).

The internal energy motion associated with the spin degree of freedom can be imagined as a system of infinitesimal vortex cells (loops) that fill the field volume, each cell carrying the circulation proportional to $s_3(\mathbf{r})$ [11–13] (see figure 1). The adjacent loops normally cancel each other, and only in areas where function $s_3(\mathbf{r})$ is not homogeneous, a certain residuary energy flow results in the non-zero SMD (figure 1b) [5,7]

$$\mathbf{p}_S = \frac{1}{2\omega c}\left(\mathbf{e}_x \frac{\partial s_3}{\partial y} - \mathbf{e}_y \frac{\partial s_3}{\partial x}\right) = -\frac{1}{2\omega c}\left[\mathbf{e}_z \times \nabla_\perp s_3\right]. \tag{7}$$

This residuary flow is concentrated in the regions of strong inhomogeneity of $s_3(\mathbf{r})$, e.g., at the boundary of a transversely limited circularly polarized beam [11,12] and is a source of the spin angular momentum of light [7]. This intuitive picture of the spin flow origination can be generalized to the Schrödinger's and Dirac's probability currents to explain the electron spin in quantum mechanical systems [12–15].

The field momentum decomposition (4) and (5) has recently been recognized to play an important role for the general electromagnetic theory [16–18]. The 3D version of equation (7),

$$\mathbf{p}_S = \frac{1}{2\omega c}\nabla \times \mathbf{\Phi}, \tag{8}$$

unites the SMD of the field with the 3D spin density $\mathbf{\Phi}$ which coincides with the field chirality flow

$$\mathbf{\Phi} = \frac{cg}{2}\operatorname{Im}\left(\frac{1}{\mu}\mathbf{E}^* \times \mathbf{E} + \frac{1}{\varepsilon}\mathbf{H}^* \times \mathbf{H}\right) \tag{9}$$

associated to the time-average chirality density $K = g\operatorname{Im}\left(\mathbf{H}^* \cdot \mathbf{E}\right)$ [9]. The optical chirality is an integral of motion in case of the dual electromagnetic symmetry, i.e. if the field equations preserve their form upon transformations

$$\mathbf{E} \to \mathbf{H}, \quad \mathbf{H} \to -\mathbf{E}, \quad \varepsilon \rightleftarrows \mu . \tag{10}$$

In this context, equations (4), (5) and (8), (9) reflect the fundamental relations between the dual symmetry, chirality conservation and spin of electromagnetic fields [19].

Additionally, the internal flow pattern provides a universal set of informative and physically meaningful parameters applicable to arbitrary light fields and suitable both for their physical characterization and for many modern applications [7,20].

All the exposed facts characterize the optical currents as well as their spin and orbital constituents as useful instruments for the study of complex optical fields, their evolution and transformations. However, their wide application is hampered by the lack of direct ways for their detection and/or measurement. Of course, as observable quantities they can be measured in experiment but the only regular available method is based on determining the field amplitudes and phases followed by the energy flow calculation via the known formulae (2) – (5).

Promising approaches to immediate measurement of the energy flow parameters can be grounded on the field mechanical action accompanied by direct observations of the probe particle motion [7,10,20–28]. Their detective abilities are based on the rapidly developing techniques for the particle trapping and micromanipulation by optical fields [20] and have been demonstrated in a long series of experiments (some recent data can be found, e.g., in Refs. [20–23,29–31]). However, their application for the quantitative study of optical currents requires detailed investigations of

(i) the physical mechanisms and peculiar features inherent in the mechanical action of the internal energy flows and their spin and orbital constituents,

(ii) characteristic details distinguishing them from some non-Poynting sources of the optical field mechanical action, e.g., the gradient force [20].

To this purpose, the known experimental demonstrations obtained in the real situations of strongly focused (for the OMD manifestation; see, e.g., Refs. [29–31]) or strongly inhomogeneous (for demonstration of the SMD-induced forces [21,22]) circularly polarized beams have been supplemented by numerical studies of the optically-induced mechanical actions emerging in relatively simple configurations of optical fields [10,32]. Many research works have shown that the forces that can be naturally associated with the spin and orbital currents really exist and can be measured; however, their physical nature and regularities they obey are still questionable.

It seems well established that the mechanical action of the orbital flow can be treated as a sort of local light pressure: a particle partially absorbs, reflects or deflects the transverse field momentum and experiences the recoil force that causes the translational (locally) and/or orbital (globally, regarding the field configuration) motion. Actually, if a small particle is weakly coupled to the optical field (the interaction cross section is much less than the wavelength and the characteristic scale of the field inhomogeneity), the observed trajectory of a particle performs the quantum weak measurement of the field orbital momentum, and the current particle location at, say, $\mathbf{R} = \mathbf{R}_0$ performs post-selection of the photon state with well defined coordinate $\mathbf{R}_0$ [17,33].

In contrast to the apparent clarity and even multiplicity of interpretations of the OMD-induced mechanical action, the mechanism by which the spin flow can move the particles remains unclear. A mere absorption or reflection of the "inhomogeneous rotation" outlined by figure 1b can obviously cause the spinning motion (which is thus considered as a characteristic manifestation of the spin momentum [20,30,31]) but no translation or orbital motion. So the questions appear about the physical nature of the SMD-induced translation and on discrepancies or similarities between the spin-induced mechanical action and the usual light pressure or the gradient force effects. The existing data give little ground for the definite answer, and the main reason is that the experiments [21,22] and corresponding calculations [32,10] were performed with relatively large particles (of the near-wavelength and larger size) for which the light-scattering effects, including the field-induced ponderomotive action, are strongly mediated by the particle size and optical properties. In this case, the field-particle interaction involves many orders of the multipole expansion expressed by the corresponding terms of the Mie-theory series [34]. As a result, the observed mechanical action depends rather intricately on the particle radius, showing apparently irregular oscillations, and the forces that can be ascribed to the SMD or OMD carry no distinct features which could be treated as their characteristic attributes. The absolute values of these forces are also quite commensurable, though each of them may be extremely small and even vanish at some special conditions [10,32] whose physical nature is not clear.

However, for very small particles, whose radius $a$ satisfies the Rayleigh scattering condition

$$ka \ll 1 \tag{11}$$

where

$$k = \frac{\omega}{c} n$$

is the radiation wavenumber in the medium with the refraction index $n=\sqrt{\varepsilon\mu}$, the light–particle interaction appears to be free from such complications and the absolute values of all optical force constituents $\mathbf{F}^{(j)}$ behave according to simple power-law rules [10,32]

$$\log\left|\mathbf{F}^{(j)}\right| = \nu_j \log\left(ka\right) + \text{const}. \tag{12}$$

Integer numbers $\nu_j$, depending on the particle properties, are the determinative signatures of these rules, distinctly different for forces caused by different 'ponderomotive factors' conventionally denoted by index *j*: the energy gradient (*j* = *G*), the OMD (*j* = *O*) or the SMD (*j* = *S*). Dependence (12) is remarkably simple, and one can expect that it is related to the physical nature and specific characters of the different optical force constituents in the most direct and clear manner. In the rest of the paper, we inspect how the numbers $\nu_j$ can serve the characteristic marks for these relations and how the optical forces exerted on the Rayleigh-scattering particles can help to disclose and to classify the physical mechanisms by which the different sources of the field mechanical action manifest themselves.

## 2. Analytical calculation of the optical force

For the purposes of this study, an appropriate analytical model of the field-induced mechanical action would be helpful, even if it would not supply exact numerical results. Such an approximate model applicable to small particles satisfying the condition (11) has been recently developed. It is the dipole model of the field-particle interaction, primarily built for the electric-dipole particles [23–25] and extended to its accomplished form by including the magnetic particle properties [26–28].

Generally, in the dipole approximation the total force experienced by a spherical particle placed within an electromagnetic field can be represented as a sum of three terms (see equations (42) – (44) of Ref. [26])

$$\mathbf{F} = \mathbf{F}_e + \mathbf{F}_m + \mathbf{F}_{em}. \tag{13}$$

The first summand expresses the electric dipole force; after light modification of equation (42) of Ref. [26] it reads

$$\mathbf{F}_e = \frac{1}{4}\text{Re}\left(\alpha_e\right)\nabla\left|\mathbf{E}\right|^2 + \frac{1}{2}\text{Im}\left(\alpha_e\right)\left\{k\sqrt{\frac{\mu}{\varepsilon}}\text{Re}\left(\mathbf{E}^*\times\mathbf{H}\right) + \text{Im}\left[\left(\mathbf{E}^*\cdot\nabla\right)\mathbf{E}\right]\right\} \tag{14}$$

where $\alpha_e$ is the particle electric polarizability that determines the dipole moment $\mathbf{d}_e$ induced in the particle due to external electric field $\mathbf{E}$ via equation $\mathbf{d}_e = \alpha_e\mathbf{E}$. In (14), the first summand is the gradient force depending on the inhomogeneity of the field energy distribution. The rest of the expression ("scattering force" or the "radiation pressure force" [25,26]) is associated with the field momentum (2) and its separate parts (4), (5). In turn, this second contribution consists of the "dissipative radiation force" [25], or "traditional radiation pressure" [24] (first summand in the figure brackets) and the "field gradient force" [25], named also "curl force associated to the nonuniform distribution of the spin angular momentum" [24] (second summand in the figure brackets). With allowance for the Maxwell equation $\nabla\times\mathbf{E} = i\frac{\omega}{c}\mu\mathbf{H}$, the whole expression in figure brackets acquires the form

$$\text{Im}\left[\mathbf{E}^*\times\left(\nabla\times\mathbf{E}\right)+\left(\mathbf{E}^*\cdot\nabla\right)\mathbf{E}\right] = \text{Im}\left[\mathbf{E}^*\cdot\left(\nabla\right)\mathbf{E}-\left(\mathbf{E}^*\cdot\nabla\right)\mathbf{E}+\left(\mathbf{E}^*\cdot\nabla\right)\mathbf{E}\right] = \text{Im}\left[\mathbf{E}^*\cdot\left(\nabla\right)\mathbf{E}\right]$$

which in view of (1) and second equation (5) leads to the final representation for the electric dipole force

$$\mathbf{F}_e = \frac{1}{2g\varepsilon}\text{Re}\left(\alpha_e\right)\nabla w_e + \frac{\omega\mu}{g}\text{Im}\left(\alpha_e\right)\mathbf{p}_O^e. \tag{15}$$

This result is known but relation (15) explicitly shows that the field-momentum-induced contribution of the dipole force is proportional to the electric part of the OMD.

The second term of (13) represents the magnetic dipole force. Starting with equation (43) of Ref. [26], quite similarly it can be reduced to the form

$$\mathbf{F}_m = \frac{1}{2g\mu}\operatorname{Re}\left(\alpha_m\right)\nabla w_m + \frac{\omega\varepsilon}{g}\operatorname{Im}\left(\alpha_m\right)\mathbf{p}_O^m \qquad (16)$$

which differs from (15) only by replacement of the electric characteristics with the corresponding magnetic ones (see the third equation (5)). Here the magnetic polarizability $\alpha_m$ determines the induced magnetic dipole moment by relation $\mathbf{d}_m = \mu^{-1}\alpha_m\mathbf{H}$ (for the sake of equations' symmetry, this deviates a bit from the 'standard' magnetic polarizability definition in Refs. [26–28]).

The last term of (13) appears owing to interference between the electric and magnetic dipole radiation and due to formula (44) of [26] and equations (2), (3) it equals

$$\mathbf{F}_{em} = -\frac{\omega}{3g}k^3\operatorname{Re}\left(\alpha_e^*\alpha_m\right)\left(\mathbf{p}_S + \mathbf{p}_O\right) + \frac{\omega}{3c}k^3\operatorname{Im}\left(\alpha_e^*\alpha_m\right)\operatorname{Im}\left(\mathbf{E}^*\times\mathbf{H}\right). \qquad (17)$$

Note that the separate electric (15) and magnetic (16) dipole contributions are completely governed by the energy density (1) and OMD (5) parameters; the SMD-induced mechanical action emerges only in the electric-magnetic dipole interaction term (17). Now we recast the dipole force terms (13) in agreement to their physical origins. The force component appearing due to the inhomogeneous energy distribution (gradient force) is expressed by relation

$$\mathbf{F}^{(G)} = \frac{1}{2g\varepsilon}\operatorname{Re}\left(\alpha_e\right)\nabla w_e + \frac{1}{2g\mu}\operatorname{Re}\left(\alpha_m\right)\nabla w_m, \qquad (18)$$

the force associated with the field OMD (orbital flow) equals

$$\mathbf{F}^{(O)} = \frac{\omega}{g}\left\{\left[\mu\operatorname{Im}\left(\alpha_e\right) - \frac{k^3}{3}\operatorname{Re}\left(\alpha_e^*\alpha_m\right)\right]\mathbf{p}_O^e + \left[\varepsilon\operatorname{Im}\left(\alpha_m\right) - \frac{k^3}{3}\operatorname{Re}\left(\alpha_e^*\alpha_m\right)\right]\mathbf{p}_O^m\right\}, \qquad (19)$$

and the force induced by the SMD (spin flow) is

$$\mathbf{F}^{(S)} = -\frac{\omega}{3g}k^3\operatorname{Re}\left(\alpha_e^*\alpha_m\right)\mathbf{p}_S. \qquad (20)$$

Also, in the dipole approximation an additional contribution appears expressed by the last term of (16) with rather special properties whose detailed analysis will be given further (see (35)):

$$\mathbf{F}^{(a)} = \frac{\omega}{3g}k^3\operatorname{Im}\left(\alpha_e^*\alpha_m\right)\mathbf{p}_a, \quad \mathbf{p}_a = \frac{g}{c}\operatorname{Im}\left(\mathbf{E}^*\times\mathbf{H}\right). \qquad (21)$$

Expressions (18) – (21) show that decomposition of the energy and momentum densities performed in equations (1) and (3) – (5) acquires additional physical meaning in the mechanical-action context: the optical force exerted on a small spherical particle is affected by the partial contributions $w_e$, $w_m$, $\mathbf{p}_O^e$, $\mathbf{p}_O^m$ and $\mathbf{p}_S$ rather than by the 'total' energy gradient and field momentum. The characters of these dependencies and numerical values of the force constituents are crucially determined by the electric and magnetic polarizabilities of the particle which, in turn, depend on the particle permittivity $\varepsilon_p$ and permeability $\mu_p$, both being generally complex quantities.

For example, the electric polarizability can be taken in the form [26]

$$\alpha_e = \frac{\alpha_e^0}{1 - i\dfrac{2}{3\varepsilon}k^3\alpha_e^0} \approx \alpha_e^0 + i\frac{2}{3\varepsilon}k^3\left|\alpha_e^0\right|^2, \qquad (22)$$

$$\alpha_e^0 = \varepsilon a^3\frac{\varepsilon_p - \varepsilon}{\varepsilon_p + 2\varepsilon}, \qquad (23)$$

where the $i$-proportional terms of (22) (usually rather small but of principal importance due to their imaginary character) follow as a result of the radiation friction [35]. The magnetic polarizability at optical frequencies is considered negligible in most cases; however, impressive magnetodielectric properties have recently been demonstrated and thoroughly investigated for submicrometer Si, Ge and $TiO_2$ spheres in the wavelength range 1 – 3 μm [36–38]. Allowance for the particles' magnetic polarizability enabled the authors of Refs. [26–28] to describe a number of new peculiar features associated with the field-induced mechanical action; nevertheless, their efforts were not aimed to the search of correspondence between the optical forces and the spin and orbital degrees of freedom of light.

So, based on the approach of Refs. [26–28], the magnetic polarizability can be treated similarly to the electric one which results in equations

$$\alpha_m = \frac{\alpha_m^0}{1 - i\dfrac{2}{3\mu}k^3\alpha_m^0} \approx \alpha_m^0 + i\frac{2}{3\mu}k^3\left|\alpha_m^0\right|^2, \tag{24}$$

$$\alpha_m^0 = \mu a^3 \frac{\mu_p - \mu}{\mu_p + 2\mu}. \tag{25}$$

Note that the polarizabilities are not determined by the absolute electric or magnetic properties of the particle but rather by their relative values with respect to the permittivity and permeability of the medium; in this context, a particle can show, say, magnetic properties being 'physically' non-magnetic, and vice versa. In what follows we will continue this 'operational' usage of terms calling a particle 'non-electric' or 'non-magnetic' if its basic polarizability (23) or (25), respectively, is zero.

An important consequence of equations (22) – (25) is that the real and imaginary parts of the polarizabilities strictly obey the power-law dependences on $a$ which, due to (18) – (21), justifies the 'empirical' rule (12) and enables one to classify the characteristic indices $\nu_j$ with account for the particle electric and magnetic properties. For simplicity and to keep the dipole model applicable, we avoid the plasmonic resonance phenomena [39] and exclude situations where $\mu_p \approx -2\mu$ or $\varepsilon_p \approx -2\varepsilon$; then the main conclusions on the character of the dependence (12) can be derived merely from the observation that the 'basic' polarizabilities $\alpha_e^0$ and $\alpha_m^0$ scale up with $a^3$ and the additive imaginary correction terms in (22), (24) are proportional to $\left|\alpha_{e,m}^0\right|^2 \sim a^6$. The conclusions depend on whether the quantities $\alpha_e^0$ and $\alpha_m^0$ are complex (absorptive particles, $\mathrm{Im}\left(\alpha_{e,m}^0\right) \neq 0$) or real (non-absorptive particles, $\mathrm{Im}\left(\alpha_{e,m}^0\right) = 0$).

a) The gradient force (18) is completely determined by real parts of the 'basic' polarizabilities (23), (25). Therefore, it exists whenever the particle shows any magnetic or electric properties differing from those of the ambient medium, and in all cases it grows proportionally to $a^3$ (see Table 1).

b) In contrast, the OMD-induced force (19) is governed by the imaginary parts of the electric and magnetic polarizabilities. If the particle normally absorbs radiation, at least one of the quantities $\alpha_e^0$ and $\alpha_m^0$ satisfies the condition $\mathrm{Im}\left(\alpha_{e,m}^0\right) > 0$ and the leading terms of the force expression (19),

$$\mathbf{F}^{(O)} = \frac{\omega}{g}\left[\mu \,\mathrm{Im}\left(\alpha_e\right)\mathbf{p}_O^e + \varepsilon \,\mathrm{Im}\left(\alpha_m\right)\mathbf{p}_O^m\right], \tag{26}$$

behave as $a^3$ (rows 3, 4 and 6 of Table 1). Otherwise, for non-absorptive particles, (22) – (25) yield

$$\mathrm{Im}\left(\alpha_e\right) = \frac{2k^3}{3\varepsilon}\left(\alpha_e^0\right)^2, \quad \mathrm{Im}\left(\alpha_m\right) = \frac{2k^3}{3\mu}\left(\alpha_m^0\right)^2, \tag{27}$$

and (19) can be reduced to

$$\mathbf{F}^{(O)} = \frac{\omega}{g}\frac{k^3}{3}\left\{\left[2\frac{\mu}{\varepsilon}\left(\alpha_e^0\right)^2 - \alpha_e^0\alpha_m^0\right]\mathbf{p}_O^e + \left[2\frac{\varepsilon}{\mu}\left(\alpha_m^0\right)^2 - \alpha_e^0\alpha_m^0\right]\mathbf{p}_O^m\right\}, \tag{28}$$

which is proportional to $a^6$ (rows 1, 2 and 5 of Table 1).

**Table 1.** Indices of the power-law dependence (12) for the optical forces of different origins. Conventionally, the particle is termed non-magnetic (non-electric) if its permeability (permittivity) does not differ from that of the ambient medium; this coincides with the usual definitions if the medium is vacuum.

| No | Sort of particle | Index of power (coefficient $\nu_j$ in (12)) and the lowest multipole order: Gradient force (18), $\nu_G$ | OMD force (19), $\nu_O$ | SMD force (20), $\nu_S$ |
|---|---|---|---|---|
| 1 | Non-magnetic non-absorptive<br>$\alpha_m^0 = 0$, $\alpha_e^0 \neq 0$, $\mathrm{Im}\left(\alpha_e^0\right) = 0$ | 3<br>electric dipole | 6<br>electric dipole | 8<br>electric quadrupole |
| 2 | Non-electric magnetic non-absorptive<br>$\alpha_e^0 = 0$, $\alpha_m^0 \neq 0$, $\mathrm{Im}\left(\alpha_m^0\right) = 0$ | 3<br>magnetic dipole | 6<br>magnetic dipole | 8<br>magnetic quadrupole |
| 3 | Non-magnetic absorptive<br>$\alpha_m^0 = 0$, $\alpha_e^0 \neq 0$, $\mathrm{Im}\left(\alpha_e^0\right) \neq 0$ | 3<br>electric dipole | 3<br>electric dipole | 8<br>electric quadrupole |
| 4 | Non-electric magnetic absorptive<br>$\alpha_e^0 = 0$, $\alpha_m^0 \neq 0$, $\mathrm{Im}\left(\alpha_m^0\right) \neq 0$ | 3<br>magnetic dipole | 3<br>magnetic dipole | 8<br>magnetic quadrupole |
| 5 | Magnetic dielectric non-absorptive<br>$\alpha_e^0 \neq 0$, $\alpha_m^0 \neq 0$, $\mathrm{Im}\left(\alpha_m^0\right) = \mathrm{Im}\left(\alpha_e^0\right) = 0$ | 3<br>electric & magnetic dipoles separately | 6<br>electric & magnetic dipoles separately | 6<br>interaction of electric & magnetic dipoles |
| 6 | Magnetic electric absorptive<br>$\alpha_e^0 \neq 0$, $\alpha_m^0 \neq 0$; at least one of the quantities $\mathrm{Im}\left(\alpha_e^0\right)$, $\mathrm{Im}\left(\alpha_m^0\right)$ is non-zero | 3<br>electric & magnetic dipoles separately | 3<br>electric & magnetic dipoles separately | 6<br>interaction of electric & magnetic dipoles |

The electric and magnetic contributions of the OMD force (summands in the figure brackets of (26) and (28)) are always directed along the local vectors $\mathbf{p}_O^e$ and $\mathbf{p}_O^m$, respectively. Consequently, in general case $\mathbf{p}_O^e \neq \mathbf{p}_O^m$, direction of the total OMD force deviates from the OMD direction since $\mathbf{p}_O^e$ and $\mathbf{p}_O^m$ enter expressions (26) and (27) with different coefficients which play roles of the force 'sensitivities' to the electric and magnetic OMD constituents.

In many usual situations, electric and magnetic parts of the OMD are equal, at least approximately (e.g., in paraxial beams) but their contributions to the total OMD force may differ strongly because of the coefficients standing before $\mathbf{p}_O^e$ and $\mathbf{p}_O^m$ in (26) and (28). For example, a non-magnetic (non-electric) – with respect to the ambient medium, see the note below (25) – particle does not feel the magnetic (electric) part of the OMD. However, in any such case the total OMD force is proportional to the OMD $\mathbf{p}_O = 2\mathbf{p}_O^e = 2\mathbf{p}_O^m$ and $\mathbf{F}^{(O)}$ is parallel to $\mathbf{p}_O$. Regardless of the field paraxiality, the strict proportionality between $\mathbf{F}^{(O)}$ and $\mathbf{p}_O$ is realized in situations of the dual symmetry [16,17] ("first Kerker condition" [27,40])

$$\frac{\varepsilon_p}{\varepsilon} = \frac{\mu_p}{\mu} \tag{29}$$

when

$$\mathbf{F}^{(O)} = \frac{\omega}{3g}\varepsilon\mu k^3 a^6 \left(\frac{\varepsilon_p - \varepsilon}{\varepsilon_p + 2\varepsilon}\right)^2 \mathbf{p}_O. \tag{30}$$

c) The spin-flow force (20) in the dipole approximation is proportional to the quantity

$$\mathrm{Re}\left(\alpha_e^* \alpha_m\right) \approx \mathrm{Re}\left(\alpha_e^0\right)\mathrm{Re}\left(\alpha_m^0\right) + \mathrm{Im}\left(\alpha_e^0\right)\mathrm{Im}\left(\alpha_m^0\right) \tag{31}$$

and varies as $a^6$ (see (23), (25) and rows 5 and 6 of Table 1)). If the particle is non-absorptive, the second summand of (31) vanishes and

$$\mathbf{F}^{(S)} = -\frac{\omega}{3g}\varepsilon\mu k^3 a^6 \frac{\varepsilon_p - \varepsilon}{\varepsilon_p + 2\varepsilon}\frac{\mu_p - \mu}{\mu_p + 2\mu}\mathbf{p}_S. \tag{32}$$

Note that under the usual conditions $\mu_p > \mu$, $\varepsilon_p > \varepsilon$, $\mathbf{F}^{(S)}$ is directed oppositely to the SMD itself: in contrast to the above-considered orbital flow, the spin flow can "pull" the particle back rather than push it forward (this agrees with results of the numerical studies, though performed for non-magnetic particles [10,32]). For absorptive particles, $\mathrm{Re}\,\alpha_e^0$ and $\mathrm{Re}\,\alpha_m^0$ can be negative and then the sign of $\mathbf{F}^{(S)}$ is determined by interplay of different terms in equation (31) thus allowing $\mathbf{F}^{(S)}$ to be directed positively (along $\mathbf{p}_S$).

Interestingly, under the dual-symmetry conditions (29) (which implies the particle to be non-absorptive), the total force owing to the field momentum equals

$$\mathbf{F}^{(O)} + \mathbf{F}^{(S)} = \frac{\omega}{3g}\varepsilon\mu k^3 a^6 \left(\frac{\varepsilon_p - \varepsilon}{\varepsilon_p + 2\varepsilon}\right)^2 \left(\mathbf{p}_O - \mathbf{p}_S\right). \tag{33}$$

This means that a dual-symmetric dipole particle "feels" the orbital and spin flows with the same sensitivity but of opposite signs. While the total Poynting vector is additive, $\mathbf{p} = \mathbf{p}_O + \mathbf{p}_S$, the corresponding force is proportional to the difference $\mathbf{p}_O - \mathbf{p}_S$. This may cause paradoxical situations where the total momentum density grows due to addition of the spin and orbital currents (cf., e.g., figures 6c, d of Ref. [7]) but the mechanical action of this momentum degrades, and vice versa. In a field with equal spin and orbital momenta (of course, this may only be realized for the transverse orbital momentum components), the total momentum-induced force (33) vanishes.

d) We finalize the comparative analysis of the different dipole force contributions by briefly commenting the still non-attributed term (21). For non-absorptive particles, $\alpha_e^0$ and $\alpha_m^0$ are real and, in view of (22) – (25),

$$\mathrm{Im}\left(\alpha_e \alpha_m^*\right) \approx \frac{2k^3}{3}\alpha_e^0\alpha_m^0 \left(\frac{\alpha_e^0}{\varepsilon} - \frac{\alpha_m^0}{\mu}\right) \sim (ka)^3 \alpha_e^0 \alpha_m^0. \tag{34}$$

Comparing (34) and (21) with (28) and (20), (32) and omitting the common pre-factor $\omega k^3/3g$, we conclude that in this case $\mathbf{F}^{(a)} \sim (ka)^3 \mathbf{F}^{(O)}$ and varies as $a^9$. In the opposite situation of absorptive particles, when $\mathrm{Im}\,\alpha_{e,m}^0 \neq 0$ and $\alpha_{e,m}^0 \sim a^3$, $\mathrm{Im}\left(\alpha_e \alpha_m^*\right)$ scales up with $a^6$. Correspondingly, the force

(21) behaves as $\mathbf{F}^{(a)} \sim (\omega/3g)(ka)^3 a^3$ while the leading terms in (19) are scaled by multiplier $(\omega/g)a^3$ Therefore, for commensurable magnitudes of $\mathbf{p}_O$, $\mathbf{p}_S$ and $\mathbf{p}_a$ the force $\mathbf{F}^{(a)}$ is generally three orders in ($ka$) less than the leading momentum-induced terms (OMD action in case of absorptive particles and the combined OMD and SMD action in case of non-absorptive particles). However, this force can become significant in some special conditions aimed at studying the SMD-induced action, where the orbital momentum vanishes or is geometrically isolated (see, e.g. [32]). Then $\mathbf{F}^{(a)}$, scaled up as $a^6$ for absorptive and as $a^9$ for non-absorptive particles, may compete with the SMD-induced force that is proportional to $a^6$ or $a^8$ (see Table 1).

However, in many such situations the 'pure' field-momentum induced mechanical action can be separated from the $\mathbf{F}^{(a)}$ contribution due to specific spatial and polarization properties of the multiplier $\mathbf{p}_a$ of (21). Indeed, it can be represented in form [26]

$$\mathbf{p}_a = \frac{g}{\omega\mu}\left\{-\frac{1}{2}\nabla|\mathbf{E}|^2 + \mathrm{Re}\left[\left(\mathbf{E}^*\cdot\nabla\right)\mathbf{E}\right]\right\} = \frac{g}{\omega\varepsilon}\left\{\frac{1}{2}\nabla|\mathbf{H}|^2 - \mathrm{Re}\left[\left(\mathbf{H}^*\cdot\nabla\right)\mathbf{H}\right]\right\}. \quad (35)$$

Apparently it contains the gradient-like contribution that can be essential in some strongly ingomogeneous fields (e.g., in the field model of Ref. [28] where **E** is $z$-polarized but only depends on $x$ and $y$, the second term in braces of the first equality (35) vanishes, and the whole term behaves as the electric gradient force). However, contrary to the 'genuine' gradient force (18), the magnetic and electric field inhomogeneities act differently (compare the two parts of (35)); additionally, this gradient-like contribution appears due to a rather special combination of the electric and magnetic properties of the particle providing $\mathrm{Im}\left(\alpha_e\alpha_m^*\right) \neq 0$. In other experiments aimed to the internal energy flow detection one frequently deals with nearly plane-wave fields (beams), where $|\mathbf{H}| \approx \sqrt{\varepsilon/\mu}\,|\mathbf{E}|$, and the contribution (35) tends to zero. In particular, in a paraxial beam propagating along axis $z$, quantity (35) can be easily calculated using the first-order paraxial representations of the vectors **E** and **H** [5,7]. Since this contribution seems to be of secondary importance for our purpose of studying the mechanical manifestations of the spin and orbital optical currents, we do not reproduce the details and only formulate a preliminary conclusion that for paraxial beams vector $\mathbf{p}_a$ lies in the ($x$, $y$) plane (is completely transverse) and its absolute value can be commensurable, by the order of magnitude, with $\mathbf{p}_{O\perp}$ and $\mathbf{p}_S$. However, its dependence on the polarization and spatial distributions is rather different and reminds what was qualified in Ref. [32] as "polarization-dependent dipole force". For example, in circularly polarized paraxial beams where the SMD-induced force is expected to be the best observable, quantity (35) vanishes.

In general, properties and possible effects associated with the force constituent (21) deserves the special consideration, and this work is already in progress [27,28]. Here we only stress that, for the problem of the energy flow detection, the force $\mathbf{F}^{(a)}$ can be a source of noise, and suggest that it might be this term that is responsible for the specific polarization-dependent mechanical action that exists in a spatially inhomogeneous field with arbitrary linear polarization [32].

## 3. Spin-flow mechanical action beyond the dipole approximation

The analytical model described in the previous section predicts a zero result for the SMD-induced force exerted on non-magnetic or non-electric (in the meaning of Table 1) particles and gives therefore no information for rows 1 – 4 of the last column in Table 1. Thus the corresponding values of $\nu_S$ as well as the order of the multipole interaction responsible for the SMD-induced force cannot be obtained analytically. This motivates our resort to the numerical model used before [10,32] (figure 2) which represents, maybe, the simplest non-trivial example of an optical field with distinctly expressed spin energy flow. Here the spatially inhomogeneous optical field is composed of two circularly polarized plane waves that approach the observation plane $z = 0$ under symmetric angles $\gamma$ and $-\gamma$ both lying in plane $x = 0$.

If both waves have the same amplitude $E_0$, in the plane $z = 0$ a standing-wave structure is formed with the internal energy flow of spin nature directed along axis $x$ whereas the internal (transverse) orbital flow is absent. Under paraxial condition $\gamma << 1$ the SMD distribution obeys the relation [10]

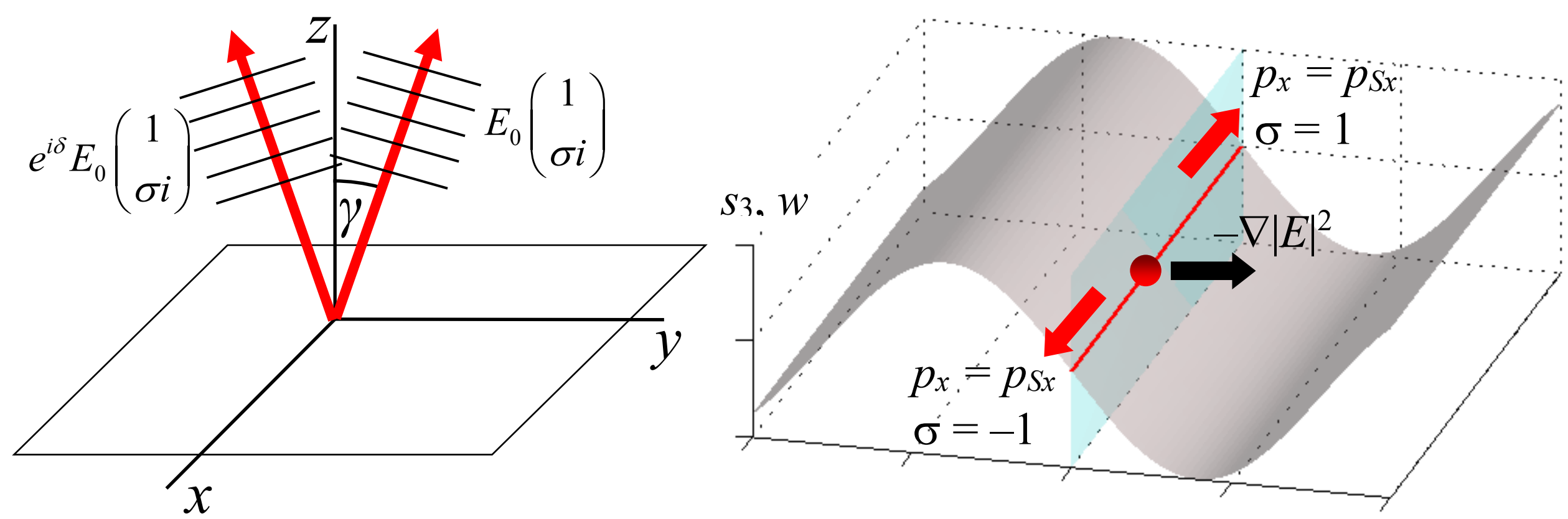

**Figure 2**. (a) Two-plane-wave model of the spatially inhomogeneous optical field with non-uniform polarization [10]; both waves possess identical circular polarization and approach the observation plane $z = 0$ symmetrically with the angle of incidence $\gamma$, and differ by the initial phase shift $\delta$. (b) Energy distribution over the plane $z = 0$: the transverse energy flow of spin nature $p_{Sx}$ is directed along or against axis $x$, regarding the polarization helicity $\sigma$, the gradient force is parallel to axis $y$.

$$p_{Sx} = \frac{4g}{c}\sqrt{\frac{\varepsilon}{\mu}}\,\sigma\gamma\left|E_0\right|^2\sin\left(\delta - 2\gamma ky\right) \tag{36}$$

where $\delta$ is the phase difference between the two plane waves, $\sigma = \pm 1$ denotes the polarization handedness.

The force exerted on a particle is calculated via the following scheme [10]. Due to the particle presence, the scattered field $\mathbf{E}_{sc}$, $\mathbf{H}_{sc}$ emerges that adds to the incident field $\mathbf{E}$, $\mathbf{H}$ [34] causing the total field momentum density (2), (3) to be changed by

$$\Delta\mathbf{p} = \frac{g}{c}\mathrm{Re}\left[\left(\mathbf{E}^* + \mathbf{E}^*_{sc}\right)\times\left(\mathbf{H} + \mathbf{H}_{sc}\right) - \mathbf{E}^*\times\mathbf{H}\right] = \frac{g}{c}\mathrm{Re}\left(\mathbf{E}^*_{sc}\times\mathbf{H}_{sc} + \mathbf{E}^*\times\mathbf{H}_{sc} + \mathbf{E}^*_{sc}\times\mathbf{H}\right). \tag{37}$$

The change of the field momentum results in a recoil force acting on the particle. This force can be determined by deriving the field momentum flux through the spherical surface $A_R$ with radius $R \to \infty$ surrounding the particle:

$$\mathbf{F} = -cn\oint_{A_R}\Delta\mathbf{p}\,dA = -cnR^2\oint\Delta\mathbf{p}\,d\Omega \tag{38}$$

where $d\Omega$ indicates integration over the solid angle. Note that the recoil force is determined by the change of the Minkowski momentum $n^2\Delta\mathbf{p}$ [3,4], which includes the reaction of the medium, rather than by the 'pure' Abraham momentum implied in (37) that was mistakenly used in equation (6) of [10].

The scattered field is linearly related to the incident field via the scattering matrix, $\mathbf{E}_{sc} = \mathbf{S}\mathbf{E}$, that is calculated for each plane wave independently by means of the standard Mie theory [34]. In this theory, the elements of $\mathbf{S}$ are represented by infinite series with coefficients $a_j$ and $b_j$ ($0 < j <$

∞) which can be considered the electric and magnetic moments of multipole sources induced due to the field–particle interaction [40]. To attribute the force (38) components to certain multipole orders, we calculated the force approximately, taking into account some first terms of the series, and compared the results with the exact data. The SMD-induced force can be easily identified with the $x$-component of the total recoil force (38), $\mathbf{F}^{(S)} = \mathbf{e}_x F_x$, as the spin flow (36) is the only $x$-directed factor of the incident field, and only the $F_x$ component changes the sign together with the polarization handedness $\sigma$.

The results of the SMD force calculation for particles corresponding to rows 1 – 4 of Table 1 are presented in Table 2. For convenience, the numerical values are normalized via division by the incident Minkowski momentum flux through the particle cross section

$$P_0 = 4g\varepsilon \left|E_0\right|^2 \cdot \pi a^2 .$$

Together with the data for non-electric and non-magnetic particles, in the last column of Table 2 the results obtained for a particle with both electric and magnetic non-zero polarizabilities (row 5 of Table 1) are also presented. They allow one to compare ‘qualities’ of the different approximations and the accuracy of the dipole model of section 2. In particular, close similarity of the data in rows 1 – 6 of the last column of Table 2 testifies for the validity of the dipole model and corresponding analytical expressions for the SMD-induced force in case of small magnetic dielectric particles.

**Table 2**. Normalized SMD-induced optical force calculated for the field model of figure 2 with $\sigma$ = 1, $\gamma$ = 0.01 upon conditions $\varepsilon$ = 1, $\mu$ = 1, $ka$ = 0.1: comparison of the data obtained by different approximations based on the Mie expansion for the scattered field, and via the analytic model of section 2. The particle parameters are illustrative.

| No | Method of calculation | | Particle parameters | | | | |
|---|---|---|---|---|---|---|---|
| | | | $\mu_p$ = 1.5; $\varepsilon_p$ = 1 non-electric, magnetic, non-absorptive | $\mu_p$ = 1.5+2i; $\varepsilon_p$ = 1 non-electric, magnetic, absorptive | $\mu_p$ = 1; $\varepsilon_p$ = 1.5 electric, non-magnetic, non-absorptive | $\mu_p$ = 1; $\varepsilon_p$ = 1.5+2i electric, non-magnetic, absorptive | $\mu_p$ = 1.5; $\varepsilon_p$ = 1.5 dielectric, magnetic |
| | 1 | | 2 | 3 | 4 | 5 | 6 |
| 1 | Non-zero coefficients in the scattered field expansion (see [34]) | All (exact result) | $-3.319\cdot10^{-11}$ | $-4.452\cdot10^{-10}$ | $-3.319\cdot10^{-11}$ | $-4.452\cdot10^{-10}$ | $-5.704\cdot10^{-8}$ |
| 2 | | $a_1$, $b_1$, $a_2$, $b_2$ | $-3.319\cdot10^{-11}$ | $-4.452\cdot10^{-10}$ | $-3.319\cdot10^{-11}$ | $-4.452\cdot10^{-10}$ | $-5.704\cdot10^{-8}$ |
| 3 | | $a_1$, $b_1$, $a_2$ | $-6.636\cdot10^{-11}$ | $-8.523\cdot10^{-10}$ | $-3.319\cdot10^{-11}$ | $-4.452\cdot10^{-10}$ | $-5.707\cdot10^{-8}$ |
| 4 | | $a_1$, $b_1$, $b_2$ | $-3.319\cdot10^{-11}$ | $-4.452\cdot10^{-10}$ | $-6.636\cdot10^{-11}$ | $-8.523\cdot10^{-10}$ | $-5.707\cdot10^{-8}$ |
| 5 | | $a_1$, $b_1$ | $-6.636\cdot10^{-11}$ | $-8.523\cdot10^{-10}$ | $-6.636\cdot10^{-11}$ | $-8.523\cdot10^{-10}$ | $-5.711\cdot10^{-8}$ |
| 6 | | $a_1$, $a_2$ | $2.212\cdot10^{-17}$ | $3.760\cdot10^{-16}$ | $3.317\cdot10^{-11}$ | $4.071\cdot10^{-10}$ | $3.327\cdot10^{-11}$ |
| 7 | | $b_1$, $b_2$ | $3.317\cdot10^{-11}$ | $4.071\cdot10^{-10}$ | $2.212\cdot10^{-17}$ | $3.760\cdot10^{-16}$ | $3.327\cdot10^{-11}$ |
| 8 | Analytic approximation (20) | | 0 | | 0 | | $-5.699\cdot10^{-8}$ |
| 9 | Coefficient $\nu_S$ in (12) – exponent in the power-law dependence $\left|\mathbf{F}^{(S)}\right| \sim (ka)^{\nu_S}$ | | 8 | | 8 | | 6 |

The data of other columns help to understand why the dipole model is not applicable for non-magnetic or non-electric particles. No matter absorbs such a particle the radiation or not, the force is calculated correctly only if the quadrupole terms of the scattered field expansion (coefficients $a_2$ and $b_2$) are taken into account. Quite expectedly, for non-magnetic particles ($\mu_p = \mu$, columns 4 and 5 of Table 2) allowance for the electric quadrupole moment ($a_2$, rows 2 and 3) is sufficient, while for non-electric particles ($\varepsilon_p = \varepsilon$, columns 2 and 3), the magnetic quadrupole moment is only important ($b_2$, rows 2 and 4). Also, both the electric and magnetic dipole contributions (coefficients $a_1$ and $b_1$) are indispensable: omitting one of these coefficients leads to essential inaccuracy even for particles with zero electric (23) or magnetic (25) polarizability (rows 6 and 7).

The reasons of such behaviour and the corresponding index $\nu_S$ of the force growth with *ka* (see (12)) can be evaluated via the following deduction. Due to (37) and (38), the recoil force is a quadratic quantity with respect to the scattered field amplitudes, and separate terms of its expansion in powers of *ka* originate from products of corresponding terms of the amplitudes' expansions. Forms of the leading terms of the field amplitude expansion are dictated by approximate expressions for the scattering coefficients valid at $ka \ll 1$,

$$a_1 = -\frac{2i}{3}(ka)^3 \frac{\varepsilon_p - \varepsilon}{\varepsilon_p + 2\varepsilon} - \frac{i}{5}(ka)^5 \frac{\varepsilon_p^2 + \varepsilon_p \varepsilon \left(m^2 - 6\right) + 4\varepsilon^2}{\left(\varepsilon_p + 2\varepsilon\right)^2},$$

$$b_1 = -\frac{2i}{3}(ka)^3 \frac{\mu_p - \mu}{\mu_p + 2\mu} - \frac{i}{5}(ka)^5 \frac{\mu_p^2 + \mu_p \mu \left(m^2 - 6\right) + 4\mu^2}{\left(\mu_p + 2\mu\right)^2}, \tag{39}$$

$$a_2 = -\frac{i}{15}(ka)^5 \frac{\varepsilon_p - \varepsilon}{2\varepsilon_p + 3\varepsilon}, \quad b_2 = -\frac{i}{15}(ka)^5 \frac{\mu_p - \mu}{2\mu_p + 3\mu} \tag{40}$$

($m = \sqrt{\varepsilon_p \mu_p / \varepsilon\mu}$ is the relative refraction index of the particle), which can be readily derived from the known formulae for $a_j$ and $b_j$ [34,40]. The lowest-order contributions associated with the quadrupole terms $a_2$ and $b_2$ may stem from products $a_1 a_2$, $a_1 b_2$, $b_1 a_2$ and $b_1 b_2$ which all behave as $a^8$. Similarly, if one of the particle polarizabilities (electric or magnetic) and, accordingly, the first term in one of equations (39) is zero, the electric – magnetic dipole coupling (product $a_1 b_1$) also obeys the rule $\sim a^8$. Direct numerical calculations immediately support this conclusion (see also Ref. [10]), which is reflected in the row 9 of Table 2 and in the last column of Table 1 (rows 1 – 4).

Therefore, the SMD-induced mechanical action, exerted on a non-magnetic or non-electric particle, generally consists of two contributions of approximately equal significance. The first one is essentially of the same nature as the SMD force exerted on particles with $\alpha_e^0 \neq 0$ and $\alpha_m^0 \neq 0$ and follows from interaction between the electric and magnetic dipole moments (but one of these moments is described by the second term of the appropriate equation (39) and is proportional to $a^5$). The second one is a consequence of interference between the dipole and quadrupole contributions (40) of the scattered field (electric dipole and quadrupole for particle with $\alpha_m^0 = 0$ and magnetic ones if $\alpha_e^0 = 0$).

## 4. Conclusion

In this paper, the optical force acting on a spherical particle with radius satisfying the Rayleigh scattering condition $ka \ll 1$ has been analyzed. In the dipole approximation, a completely analytical approach is possible which permits us to establish the main regularities relating different sources of the ponderomotive action of electromagnetic fields ('ponderomotive factors'): the energy density gradient and the energy flow (field momentum), in particular, its orbital and spin parts. Noteworthy, in view of the mechanical effects, the field momentum division into the spin and orbital constituents acquires additional physical meaning because of the difference in mechanical actions associated with both contributions.

On the other hand, the actual force experienced by a probe particle crucially depends on the particle properties with respect to the ambient medium. In this context, all particles can be classified into 'non-magnetic' (non-zero electric polarizability (23), $\alpha_e^0 \neq 0$, and zero magnetic polarizability (25), $\alpha_m^0 = 0$), 'non-electric' (opposite conditions $\alpha_e^0 = 0$, $\alpha_m^0 \neq 0$) and electric-magnetic ones with both non-vanishing polarizabilities, $\alpha_e^0 \neq 0$, $\alpha_m^0 \neq 0$; in each class, absorptive ($\mathrm{Im}\,\alpha_{e,m}^0 \neq 0$) and non-absorptive ($\mathrm{Im}\,\alpha_{e,m}^0 = 0$) particles distinctly differ. In compliance with this division, the energy gradient, orbital and spin momenta of the field can also be represented as sums of the electric and magnetic contributions whose separate existence manifests itself via the selective reaction of particles with appropriate electromagnetic nature. Only in case when the electromagnetic dual symmetry is extended to the particle properties ('dual' particle, equation (29)), the particle motion is governed by the 'genuine' field energy gradient, orbital and spin momenta, rather than by their electric or magnetic parts (see also Ref. [17]).

In general, the analysis performed has supported previous anticipations inspired by numerical calculations [10,32]. Additionally, it explains and refines the preliminary results in some important aspects relating the peculiar features of the mechanical actions associated with different physical sources. In particular, the index of power $\nu_j$ in the power-law relation (12) between the optical force and the particle size can serve a characteristic mark of the corresponding ponderomotive factor and reflects the physical nature of its mechanical action.

It was established that in the 'pure' electric or magnetic dipole approximation, the spin flow exerts no mechanical action on sub-wavelength (Rayleigh) particles. The gradient force in such conditions grows as $a^3$, the orbital-momentum force scales up with $a^3$ for absorptive and with $a^6$ for non-absorptive particles. The lowest-order spin-momentum force is also proportional to $a^6$ and appears as a result of interference between the electric and magnetic dipole scattering if a particle possesses both dielectric and magnetic properties. For non-electric or non-magnetic particles, the spin-momentum force consists of two contributions which cannot be studied within the frame of the dipole model and were investigated by a semi-analytic approach based on the approximate representations of the Mie scattering coefficients (39), (40). The first part of the spin-momentum force follows from interaction between the electric and magnetic dipole moments (but because of the zero dipole polarizability, one of these moments only emerges in higher degrees in the particle-size expansion and is described by the second term of the appropriate equation (39)), the second one appears due to interference between the dipole and quadrupole components of the scattered field. Both contributions lead to the force scaling with $a^8$. An interesting deduction which agrees with the earlier results [10] is that the spin flow, in contrast to the orbital one, may "pull" particles against its own direction, regarding the particle electric and magnetic polarizabilities and absorption properties. In the usual case of non-magnetic particles the force directed against the spin flow is typical for dielectric particles while the conductive ones move preferably along the spin momentum. Of course, the spin-momentum-induced translation is always accompanied by the particle spinning motion that is not studied in this paper.

The above figures can make an impression that the spin-momentum force is generally much lower than the gradient and the orbital-momentum ones but this is only correct in the small-particle limit $ka \ll 1$. Numerical calculations show that when the particle size satisfies the condition $ka \gtrsim 1$, all the ponderomotive factors may perform actions of quite comparable magnitudes [10,32]. Besides, they suggest how contributions of different origins (e.g., the gradient force and the light-pressure force) can be isolated and observed separately owing to special but easily realizable field configuration: in particular, the gradient force is always orthogonal to the SMD force in the scheme of figure 2b. The results of the present paper demonstrate additional possibilities for separate observation of the mechanical actions induced by the spin and orbital momenta employing the special choice of the probe particles with necessary magnetic and/or electric properties. We hope that our results can be useful for the experimental study of optical currents and their spin and orbital constituents.

**Acknowledgments**

The author thanks Konstantin Bliokh for fruitful ideas and stimulating discussion. This work was supported, in part, by the Ministry of Education and Science of Ukraine within the frame of the budget project No 494/12 (State Registration Number 0112U001741).